\documentclass[11pt,a4paper]{article}
 \pdfoutput=1
\usepackage{jcappub}
\usepackage{amsmath}
\usepackage{amsfonts,color}
\usepackage{amssymb,float}
\usepackage{mathtools}
\usepackage[utf8]{inputenc}
\usepackage{url}
\usepackage{graphicx}
\usepackage{refstyle}
\usepackage{wasysym}
\usepackage{tabularx}
\usepackage{accents}
\usepackage{graphicx}
\usepackage{color}
\usepackage[dvipsnames]{xcolor}
\usepackage{hyperref}

\hypersetup{
    colorlinks=true,
    linkcolor=blue,
    filecolor=magenta,      
    citecolor=red
}

\usepackage{cancel} % for \cancel \Lambda
\usepackage{subfigure} % for subfigures
\usepackage{hyperref} % for hyperlinks

\makeatletter
\long\def\dddddot#1{%
  {\mathop {#1}\limits ^{\vbox to-1.4\ex@ {\kern -\tw@ \ex@ \hbox {\normalfont .....}\vss }}}%
}
\long\def\multidots#1#2{%
  \count@=0
  {{\mathop {#2}\limits ^{\vbox to-1.4\ex@ {\kern -\tw@ \ex@ \hbox {\normalfont %
  \loop%
  \ifnum#1>\count@%
  .%
  \advance\count@ by1%
  \repeat%
  }\vss }}}}%
}
\makeatother

\newcommand{\be}{\begin{equation}}
\newcommand{\ee}{\end{equation}}
\newcommand{\bs}{\begin{split}} 
\newcommand{\bea}{\begin{eqnarray}}
\newcommand{\eea}{\end{eqnarray}}

\newcommand{\mpl}{M^2_{\rm  pl}} 

\usepackage[dvipsnames]{xcolor}
\usepackage[normalem]{ulem} % for strikeout

\title{\boldmath Tadpole Cosmology: Self Tuning Without Degeneracy}

\author[a,b]{Stephen Appleby}
\author[c,1]{Reginald Christian Bernardo\note{Corresponding Author}}

\affiliation[a]{Asia Pacific Center for Theoretical Physics, Pohang 37673, Korea}
\affiliation[b]{Department of Physics, POSTECH, Pohang 37673, Korea}
\affiliation[c]{Institute of Physics, Academia Sinica, Taipei 11529, Taiwan}

\emailAdd{stephen.appleby@apctp.org}
\emailAdd{rbernardo@gate.sinica.edu.tw}

\abstract{
\textbf{
Degeneracy is a method to accommodate exact, low energy vacuum states in scalar-tensor gravitational models despite the presence of an arbitrarily large vacuum energy. However, this approach requires very particular combinations of scalar field and metric couplings in the Lagrangian. In this work we study departures from the restrictive degeneracy condition -- starting from a fiducial model containing an exact Minkowski space solution, we break the degeneracy condition in numerous simple ways to test if the resulting models maintain certain key features -- specifically the dynamical cancellation of a large vacuum energy by the scalar field and the existence of a low energy vacuum state. We highlight the role the tadpole plays in eliminating the fixed points of the dynamical system, generically rendering both the scalar field and metric time dependent. Our results indicate that when violating the degeneracy condition but preserving shift symmetry, the metric maintains an asymptotic Minkowski state, irrespective of the presence of the cosmological constant. In contrast, when shift symmetry is also broken the asymptotic behaviour can radically alter. Regardless, the non-degenerate models in this work share an attractive quality; harboring low energy, late-time asymptotic states that are independent of the vacuum energy. The tadpole allows for a broader class of non-degenerate, self-tuning models than was previously realized.
}}

\begin{document}

\maketitle
\flushbottom

%%%% start of paper %%%%

\section{Introduction}

Scalar-tensor models of gravity possess a vast phenomenology, and are widely studied within the context of inflation, dark energy, black hole phenomenology, dark matter and beyond \cite{Clifton:2011jh, Joyce:2014kja, Nojiri:2017ncd}. One interesting proposal is the existence of pseudo-vacuum solutions in which Lorentz invariance is partially broken, in such a way that the scalar degree of freedom does not relax to a constant vacuum expectation value on the background spacetime. Such states open the door to self-tuning mechanisms in which the dynamical degree(s) of freedom can cancel an arbitrary vacuum energy, leaving the metric unaffected \cite{Weinberg:1988cp, Padilla:2015aaa}. This idea was pioneered in Refs. \cite{Charmousis:2011bf, Charmousis:2011ea}, in which Minkowski space solutions were obtained despite the presence of an arbitrary vacuum energy. The scalar field equation derived from this so-called `Fab-Four' action possesses a particular structure such that it is trivially satisfied when the metric is exactly Minkowski space\footnote{More precisely, Milne spacetime.}. In this case, the scalar field remains dynamical and the Friedmann equation relates the scalar field dynamics to the vacuum energy \cite{Copeland:2012qf, Appleby:2015ysa, Babichev:2015qma, Kaloper:2013vta, Copeland:2021czt,Appleby:2012rx}. A different type of degeneracy was subsequently explored in Ref. \cite{Appleby:2018yci}, in which actions were constructed for which the scalar field and one of the Einstein equations are equivalent when the metric is de Sitter \cite{Emond:2018fvv, Appleby:2020njl, Linder:2020xey, Bernardo:2021hrz, Bernardo:2021izq, Linder:2022iqi} or Minkowski space \cite{Appleby:2020dko, Bernardo:2021bsg}. In this work we focus on the latter class of models, dubbed `well-tempering', and specifically those that admit Minkowski space vacuum solutions \cite{Appleby:2020dko, Bernardo:2021bsg}. Modern cosmology requires the existence of low energy de Sitter rather than Minkowski space, but in this work we treat Minkowski space as a useful test bed.

The requirement that an exact, static vacuum solution exists for the metric despite the presence of an arbitrarily large vacuum energy imposes stringent conditions on the form that the scalar-tensor action can possess \cite{Charmousis:2011bf, Appleby:2018yci}. Demanding that Minkowski space is a solution to the field equations regardless of the energy density of the vacuum overconstrains the dynamics, and this ansatz can only be realised if the field equations have some form of redundancy when it is imposed \cite{Appleby:2020dko}. This gives rise to a degeneracy condition, which imposes an exact relationship between different terms in the scalar field Lagrangian \cite{Appleby:2020dko, Bernardo:2021bsg}. However, fixing the Lagrangian precisely so that the model admits a flat spacetime solution can be interpreted as its own form of fine tuning, separate from the Cosmological Constant problem. It is therefore natural to question how the dynamics of this class of models changes if we relax the degeneracy condition, and what is the fate of the vacuum solutions. These questions are the focus of this work. 

We focus on the simplest model in Ref. \cite{Appleby:2020dko} that can give rise to a Minkowski space solution, and then adjust the action such that the degeneracy condition is broken whilst preserving the core features of the model -- shift symmetry and Galilean invariance. We argue that the tadpole is crucial in eliminating dynamical fixed points from the system, ensuring that even if we do not impose the degeneracy condition, the metric does not relax to the standard Cosmological Constant-driven de Sitter fixed point. Instead, the expansion rate can evolve to a Minkowski space solution asymptotically. This behaviour is quite generic for the cubic Galileon model, subject to the presence of the tadpole and shift symmetry. 

The paper will proceed as follows. In section \ref{sec:2} we introduce the action and field equations that will be used throughout this work, and elucidate the role of the tadpole in precluding the existence of de Sitter solutions. In section \ref{sec:3} we briefly review the degenerate Minkowski vacuum solutions obtained in Ref. \cite{Appleby:2020dko}. Taking a simple model as `fiducial', we relax the strict degeneracy condition in multiple ways in section \ref{sec:4}, finding that the Minkowski space solution is preserved in the asymptotic future. For balance, we include an example of breaking the degeneracy condition such that the Minkowski space solution is lost completely. We close with a discussion of our results and the future hurdles that this class of models must overcome to be considered as viable cosmological models. Related asymptotic de Sitter solutions have been recently constructed in the literature \cite{Khan:2022bxs}, which are similar in spirit to this work.

\textit{Supplementary Material.} A Mathematica notebook which can be used to reproduce the results of the paper can downloaded from \href{https://github.com/reggiebernardo/notebooks}{GitHub} \cite{reggie_bernardo_4810864}.

\textit{Conventions.} We work with the mostly-plus metric signature $(-, +, +, +)$. A dot over a variable means a derivative with respect to the cosmic time $t$ while a prime corresponds to differentiation with respect to the dimensionless time $\tau$. Subscripts on the scalar potentials $K(X)$, $V(\phi)$, $G_3(X)$, and $F(\phi)$ denote differentiation with respect to their arguments $\phi$ or $X$.

\section{Cosmology with the Tadpole}
\label{sec:2}

Throughout this work we will use the following action and corresponding field equations, obtained after imposing a flat Friedmann-Lema\^{i}tre-Robertson-Walker metric:

\begin{equation} \label{eq:action} S = \int \sqrt{-g} d^{4}x \left[ {(M_{\rm pl}^{2} + F(\phi)) R \over 2} + K(X) + V(\phi) - G_{3}(X) \Box \phi - \lambda^{3}\phi - \Lambda + {\cal L}_{\rm m} \right] \end{equation} 

\bea 
3H^2(\mpl +F)&=& \rho + \Lambda+2XK_X-K -V +6H\dot\phi XG_{3X} 
-3 F_{\phi} H\dot\phi + \lambda^{3}\phi \label{eq:fullfried}\\ 
-2\dot H\,(\mpl +F)&=& \rho + P + \ddot\phi\,( F_{\phi} -2XG_{3X}) \nonumber \\
&\qquad& \phantom{ggggg} -H\dot\phi\,(F_{\phi} -6XG_{3X})
+2XK_X + 2X F_{\phi\phi} \label{eq:fulldh}\\ 
0&=&\ddot\phi\,\left[K_X+2XK_{XX}+6H\dot\phi(G_{3X}+XG_{3XX})\right]\notag\\ 
&\qquad&+3H\dot\phi\,K_X+\lambda^3 -V_{\phi} +6XG_{3X}(\dot H+3H^2) 
-3 F_{\phi} (\dot H+2H^2)   \label{eq:fullddphi} 
\eea 

\noindent where $K(X)$ and $G_{3}(X)$ are arbitrary functions of $X = -\left( \partial_\mu \phi \right) \left( \partial^\mu \phi \right) /2$ and $F(\phi)$, $V(\phi)$ are arbitrary functions of $\phi$. Subscripts denote differentiation with respect to that variable. Due to the importance of the tadpole $\lambda^{3}\phi$ in this work, we separate it from $V(\phi)$. We have included a perfect fluid contribution ${\cal L}_{m}$ with density and pressure $\rho$, $P = w \rho$. We will initially fix $\rho = P = 0$, but keep the Cosmological Constant $\Lambda$ arbitrary and non-zero. We re-introduce matter in section \ref{sec:matter}.

When faced with a dynamical system such as ($\ref{eq:fulldh}$) and ($\ref{eq:fullddphi}$), the first step is to determine the fixed points at which the fields approach constant values. The tadpole plays a unique role in the dynamics of $\phi$ and $H$, in that it generically prevents the relaxation of the fields to Cosmological Constant-driven vacuum expectation values. To see this, we return to a simple example. We fix $F(\phi) = 0$, $V(\phi) = 0$, $G_{3}(X)=0$ and $K(X) = \epsilon X$, where $\epsilon$ is a dimensionless constant that can be rescaled to unity via a field redefinition (we decline to do so, retaining the freedom of choosing the sign of the kinetic term). The field equations are then particularly simple 

\bea 
3\mpl H^2 &=&\Lambda+\epsilon {\dot{\phi}^{2} \over 2}
 + \lambda^{3}\phi \label{eq:fried}\\ 
-2\mpl \dot H\, &=&
\,
\epsilon \,\dot{\phi}^{2} \label{eq:dh}\\ 
0&=& \epsilon \ddot\phi\, +3\epsilon H\dot\phi\, +\lambda^3
\, . \label{eq:ddphi} 
\eea 

\noindent If $\lambda = 0$, then the system admits an exact vacuum solution $3 \mpl H^{2} = \Lambda$, $\phi = 0$. When $\lambda \neq 0$ this is not a solution, and in fact it is clear that $H={\rm constant}$ is not a solution to this system. $\dot{H}=0$ implies $\dot{\phi} = 0$ which is inconsistent with the scalar field equation. Assuming there exists a solution to this system that is analytic about some $t_{0}$ that we take without loss of generality to be $t_{0}=0$, we can expand as

\begin{eqnarray}  \phi(t) &=& \sum_{n=0}^{\infty} \phi_{n}t^{n} \\
H(t) &=& \sum_{n=0}^{\infty} H_{n} t^{n} \, .
\end{eqnarray}

\noindent The field equations, expanded up to ${\cal O}(t^{2})$, are 
 
 \begin{eqnarray} \label{eq:O0fr} {\cal O}(t^{0}) \qquad : \qquad & &   3M_{\rm pl}^{2} H_{0}^{2}  = \Lambda + {\epsilon \over 2}  \phi_{1}^{2}  + \lambda^{3} \phi_{0}  \\
 \label{eq:O0dH} & & -2M_{\rm pl}^{2} H_{1}  = \epsilon  \phi_{1}^{2}  \\
  \label{eq:O0sfe} & & 2\epsilon  \phi_{2}  + 3\epsilon  H_{0}\phi_{1} + \lambda^{3} = 0 \\
 \label{eq:O1fr} {\cal O}(t) \qquad : \qquad & &    6M_{\rm pl}^{2} H_{0}H_{1}   =  2\epsilon  \phi_{1}\phi_{2}   + \lambda^{3}  \phi_{1}  \\
\label{eq:O1dH}  & & -4M_{\rm pl}^{2}  H_{2}  = 4\epsilon  \phi_{1}\phi_{2}   \\
  \label{eq:O1sfe} & & 6\epsilon \phi_{3} + 3\epsilon  \left(H_{1}\phi_{1} + 2 H_{0}\phi_{2}\right)  = 0 \\
 \label{eq:O2fr} {\cal O}(t^{2}) \qquad : \qquad & &  3M_{\rm pl}^{2} \left(H_{1}^{2} + 2H_{0}H_{2}\right) =  {\epsilon \over 2}  \left(4\phi_{2}^{2} + 6\phi_{1}\phi_{3}\right)t^{2}   + \lambda^{3} \phi_{2}  \\
 \label{eq:O2dH} & & -6M_{\rm pl}^{2} H_{3} = \epsilon  \left(4\phi_{2}^{2} + 6\phi_{1}\phi_{3}\right) \\
  \label{eq:O2sfe} & & 12\epsilon \phi_{4} + 3\epsilon  \left(H_{2}\phi_{1}  + 2H_{1}\phi_{2} \right) = 0 \, .
 \end{eqnarray} 
 
 \noindent We must specify two initial conditions out of three degrees of freedom -- $H_{0}$, $\phi_{0}$ and $\phi_{1}$, then the final one is fixed by the order ${\cal O}(t^{0})$ Friedmann equation ($\ref{eq:O0fr}$). If we search for a $H = {\rm constant}$ solution, we must ensure that $H_{i} = 0$ for all $i > 0$. The $\dot{H}$ equation ($\ref{eq:O0dH}$) forces $\phi_{1} = 0$, which from ($\ref{eq:O0sfe}$) gives $\phi_{2} = -\lambda^{3}/2\epsilon$. $H_{2}$ is zero from ($\ref{eq:O1dH}$) -- $\phi_{1} = 0$ -- but $H_{3}$ is non-zero from ($\ref{eq:O2dH}$) -- $H_{3} = -2\epsilon\phi_{2}^{2}/(3M_{\rm pl}^{2})$. The time dependence of $H_{3}$ is sourced by $\phi_{2} \sim \lambda^{3}$, which confirms that in the presence of a tadpole there is no de Sitter solution. Once we abandon the idea of an exact de Sitter vacuum solution, we can take $\phi_{1} \neq 0$ in which case the time dependence of $H$ becomes a function of a combination of $\Lambda$, $\phi_{0}$ and $\phi_{1}$. One can always fix $H_{0} = 0$ as an initial condition, but since the metric is not static this seems to be an arbitrary choice. 
 
Although we took a particularly simple example, we expect this behaviour to be generic. We will argue that static vacuum solutions are not generically present when $\lambda \neq 0$, although we discuss the generality of this statement below. The degeneracy condition considered in Ref. \cite{Appleby:2018yci} allows the metric to exist in a constant vacuum state, but the presence of the tadpole requires that the scalar field must remain dynamical. If we drop the degeneracy condition, then we can expect both the metric and scalar field to evolve.

As a counterpoint to the example in this section, cases arise where a static vacuum can be realized even when the tadpole is present. The caveat here is that these solutions cancel out the Cosmological Constant via fine tuning rather than by a dynamical mechanism. As an example, when the scalar field in the dynamical system (\ref{eq:fried}), (\ref{eq:dh}), and (\ref{eq:ddphi}) is granted a bare mass $m_\phi \neq 0$, the `constant field' solution
\begin{equation}
    \left( 3 M_\text{pl}^2 H^2 = \Lambda -  \dfrac{\lambda^6}{2m_\phi^2} , \phi = - \lambda^3/m_\phi^2 \right)
\end{equation}
is present. As expected, the tadpole contribution can eliminate the standard Cosmological Constant-driven vacuum state if $\lambda^{6} = 2\Lambda m_{\phi}^{2}$. However, for this fixed point to exist the Cosmological Constant must be cancelled by fine tuning, to produce a low energy (Minkowski) vacuum. This is a manifestation of Weinberg's no-go theorem which requires cancellation between disparate contributions to the Cosmological Constant \cite{Weinberg:1988cp, Padilla:2015aaa}. Such cases do not have the behaviour that we are searching for in this work -- a time dependent dynamical cancellation of $\Lambda$. Keeping this in mind, for the rest of the paper we focus on a model space for which degenerate vacuum solutions exist as asymptotic states.

 \section{Degenerate Field Equations}
 \label{sec:3}
 
In a series of papers \cite{Appleby:2018yci, Emond:2018fvv, Appleby:2020njl, Appleby:2020dko, Linder:2020xey, Bernardo:2021izq, Bernardo:2021bsg, Linder:2022iqi}, a set of vacuum solutions to the system of equations was obtained by imposing exact de Sitter or Minkowski ansatze then searching for models for which the ansatz is a solution. Models with such vacuum solutions must solve a degeneracy condition, which relates the scalar field potentials in the action. These vacuum states are unusual, in the sense that the metric does not respond to the scalar field dynamics or any vacuum energy present, at least at the level of the background. 
 
For example, in Ref. \cite{Appleby:2020dko} a Minkowski space solution was derived. Fixing $F(\phi) = V(\phi) = 0$ and $\rho = P = 0$, the condition for the dynamical system ($\ref{eq:fullfried}$), ($\ref{eq:fulldh}$), ($\ref{eq:fullddphi}$) to admit an exact Minkowski space solution is the degeneracy condition 
 
 \begin{equation} \label{eq:deg}   G_{3X} = -{1 \over \lambda^{3}} K_{X} \left( K_{X} + 2X K_{XX} \right) ,
 \end{equation} 
 
\noindent with additional conditions $G_{3X} \neq 0$, $K_{X} \neq 0$, $K_{X} + 2XK_{XX} \neq 0$, $\lambda \neq 0$. The simplest example can be found by taking $K(X) = \epsilon X$ and $G_{3}(X) = -\epsilon^{2}X/\lambda^{3}$. The field equations become
 
 \bea 
3\mpl H^2 &=&\Lambda + {\epsilon \over 2} \dot{\phi}^{2} - {3 \epsilon^{2} \over \lambda^{3}} H\dot\phi^{3}  + \lambda^{3}\phi \label{eq:friedt}\\ 
-2\mpl \, \dot H &=&
{ \epsilon^{2} \over \lambda^{3}}\dot\phi^{2} \ddot\phi - {3 \epsilon^{2} \over \lambda^{3}} H\dot\phi^{3}
+2\epsilon X  \label{eq:fulldht}\\ 
0&=&\ddot\phi\,\left[\epsilon - {6\epsilon^{2} \over \lambda^{3}} H\dot\phi\right]+3\epsilon H\dot\phi + \lambda^3 - {3\epsilon^{2} \over \lambda^{3}}\dot\phi^{2}(\dot H+3H^2) \, .
  \label{eq:fullddphit} 
\eea

\noindent For this system there exists an exact Minkowski space solution with $H = 0$ identically. Inserting $H = \dot{H} = 0$ into ($\ref{eq:fulldht}$) and ($\ref{eq:fullddphit}$), both equations reduce to 

\begin{equation}\label{eq:ddotp} \ddot{\phi} = -\lambda^{3}/\epsilon . \end{equation} 

\noindent This is the meaning of degeneracy -- the number of independent dynamical equations must reduce by one when we impose an ansatz that fixes one of the dynamical fields -- in this case the expansion rate $H$. By introducing dimensionless variables $\varphi = \phi/\lambda$ and $\tau = \lambda t$, the field evolves as

\begin{equation} \label{eq:phi_wtvac} \varphi = -{\tau^{2} \over 2\epsilon} + c_{1}\tau + c_{0} \end{equation}

\noindent where $c_{1}$, $c_{0}$ are dimensionless constants. The constants $c_{0}$, $c_{1}$ -- which are integration constants of the dynamics of $\phi$ -- cancel the vacuum energy in the Friedmann equation ($\ref{eq:friedt}$). The dynamical nature of the field $\varphi$ is due to the presence of the tadpole; equation ($\ref{eq:ddotp}$) is sourced by $\lambda$. We expect both the scalar field and metric to remain dynamical in this model, but the metric admits a static vacuum state due to the imposition of the degeneracy condition -- we have demanded that such a  solution exists by imposing ($\ref{eq:deg}$). 

It is natural to ask what happens if we relax the degeneracy condition. Based on our understanding of the tadpole, we expect that in the absence of a mechanism to enforce a static metric solution then both the metric and scalar field can be dynamical. For the purposes of cosmology, we are not interested in the existence of Minkowski space solutions, or even exact vacuum solutions. We expect that the standard general relativity algebraic relation $3M_{\rm pl}^{2} H^{2} = \Lambda$ is not necessarily a solution to the dynamical system when the tadpole is present, and we also know that Minkowski space is a dynamical attractor when the degeneracy equation is exactly imposed. One might hope then that violating the degeneracy condition might yield a dynamical solution for the metric as well as the scalar field in such a way that $H$ evolves towards the low energy state $H(t) \to 0$, rather than towards the general relativity $\Lambda$-driven de Sitter vacuum solution, without having to impose an exact degeneracy condition.

%%%%%%%%%%%%%% 
\begin{table*}[h!]
    \centering
\begin{tabular}{|l|c|c|c|c|}
\hline 
Model & Section & Asymptotic State & Dynamical $\cancel{\Lambda}$ & Ghost Free  \\ 
\hline 
Unequal Mass Terms $\kappa \neq \lambda$ & \ref{sec:unequal_mass} & Minkowski & \checkmark & \checkmark   \\ 
Linear Coupling $\phi R$ & \ref{sec:coupling} & Minkowski & \checkmark & \checkmark \\ 
Nonlinear Coupling $\phi^{n}R$ & \ref{sec:coupling} & Minkowski & \checkmark & \checkmark  \\ 
Including Matter $\rho \neq 0$ & \ref{sec:matter} & Minkowski & \checkmark  & \checkmark \\
Scalar Field Mass $m_{\phi}^{2}\phi^{2}$ & \ref{sec:5} & de Sitter/Unstable & \checkmark & \checkmark   \\
\hline 
\end{tabular} \\  
\caption{Summary of all models considered in this work, in which the degeneracy condition ($\ref{eq:deg})$ is broken by the addition of different terms to the scalar-tensor action. The asymptotic late time states of the dynamical system are given, as well as the ability of the scalar field to dynamically cancel the effect of $\Lambda$ and the perturbative (no-ghost) stability condition.
}
\label{tab:models} 
\end{table*}

\section{Breaking the Degeneracy Condition} 
\label{sec:4}

In this section we consider multiple ways in which the degeneracy relation might be broken. We take the exact Minkowski solution and model of the previous section as a fiducial case, then depart from it in various ways. We initially fix the matter contribution to be zero but introduce pressureless dust in section \ref{sec:matter}.

We provide a summary of the models considered in this work, and their important properties, in Table \ref{tab:models}.

\subsection{Unequal Mass Scales}
\label{sec:unequal_mass}

First we allow the mass scales associated with $G_{3}(X)$ and the tadpole to differ, and fix $V(\phi) = F(\phi) = 0$. Hence the first system that we consider has the following action 

\begin{equation}\label{eq:ac1} S = \int \sqrt{-g} d^{4}x \left[ {M_{\rm pl}^{2} R \over 2} + \epsilon X + {\epsilon^{2}  \over \kappa^{3}} X \Box \phi - \lambda^{3}\phi - \Lambda \right] \, . \end{equation} 

\noindent To admit an exact Minkowski vacuum solution we require $\lambda = \kappa$, but we do not impose that relation here. We introduce dimensionless quantities $\alpha = \lambda/\kappa$ and $\Delta = M_{\rm pl}/\lambda$, $h = H/\lambda = d\log{a}/d\tau$, $\tilde{\Lambda} = \Lambda/\lambda^{4}$ and the resulting field equations are
 \bea 
3\Delta^{2} h^2 &=& \tilde{\Lambda} + {\epsilon \over 2} (\varphi')^{2} - 3 \epsilon^{2}\alpha^{3}  h(\varphi')^{3}   + \varphi \label{eq:fr0}\\ 
-2\Delta^{2} \, h' &=&
 \epsilon^{2}\alpha^{3} (\varphi')^{2} \varphi'' - 3 \epsilon^{2} \alpha^{3} h (\varphi')^{3}
+ \epsilon (\varphi')^{2}  \label{eq:dh0}\\ 
0&=& \varphi''\,\left[\epsilon - 6\epsilon^{2}\alpha^{3} h\varphi'\right]+3\epsilon h\varphi' + 1 - 3\epsilon^{2}\alpha^{3}  (\varphi')^{2} (h'+3h^2) 
\, .  \label{eq:sf0} 
\eea 
The ansatz $h = h' =0$ is no longer a solution to this system because ($\ref{eq:dh0}$), ($\ref{eq:sf0}$) are not equivalent upon insertion of this ansatz. There are no regular fixed points to this system -- $\varphi'' = \varphi' = 0$ is not a solution to ($\ref{eq:sf0}$). There is also no $h' = 0$, $\varphi'' = 0$, $\varphi' = {\rm constant}$ solution, since the Friedmann equation is not consistent with this ansatz -- $\varphi$ would be the only time-dependent term in the equation. 

However, there exists an asymptotic solution to this system of equations such that for $\tau \gg 1$

\begin{eqnarray}  \varphi &=& \tau^{2} \sum_{n=0}^{\infty} \varphi_{n}\tau^{-n} \label{eq:phi_series} \\ 
 h &=& \tau^{-1} \sum_{n=0}^{\infty} h_{n}\tau^{-n} \, . \label{eq:h_series} 
\end{eqnarray} 

\noindent Then, order by order in a $\tau \gg 1$ expansion we have 

 \begin{eqnarray} \label{eq:O0frI} {\cal O}(h_{0}, \varphi_{0}) \qquad : \qquad & &   0 = \varphi_{0} + 2 \epsilon \varphi_{0}^{2} - 24 \epsilon^{2}\alpha^{3} h_{0}\varphi_{0}^{3}  \\
 \label{eq:O0dHI} & & 0  =  \epsilon \varphi_{0}^{2} + 2 \epsilon^{2}\alpha^{3}\varphi_{0}^{3} - 6 \epsilon^{2} \alpha^{3} h_{0} \varphi_{0}^{3}  \\
  \label{eq:O2sfeI} & & 0 =  1 + 2 \epsilon\varphi_{0} + 6\epsilon h_{0}\varphi_{0} - 12 \epsilon^{2}\alpha^{3}h_{0}\varphi_{0}^{2} - 36 \epsilon^{2}\alpha^{3}h_{0}^{2}\varphi_{0}^{2} \\ 
 \nonumber & & \\
 \label{eq:O1frI} {\cal O}(h_{1}, \varphi_{1}) \qquad : \qquad & &    0  =  \varphi_{1} - 24\epsilon^{2}\alpha^{3}h_{1} \varphi_{0}^{3} + 2 \epsilon\varphi_{0}\varphi_{1} - 36 \epsilon^{2}\alpha^{3} h_{0} \varphi_{0}^{2}\varphi_{1} \\
\label{eq:O1dHI}  & & 0  = -24\epsilon^{2}\alpha^{3}h_{1}\varphi_{0}^{3} + 4 \epsilon \varphi_{0}\varphi_{1} + 8 \epsilon^{2}\alpha^{3}\varphi_{0}^{2}\varphi_{1} - 36 \epsilon^{2}\alpha^{3} h_{0} \varphi_{0}^{2}\varphi_{1}   \\
  \label{eq:O3sfeI}  & &  0 = 6\epsilon h_{1} \varphi_{0} - 72\epsilon^{2}\alpha^{3}h_{0}h_{1}\varphi_{0}^{2} + 3 \epsilon h_{0}\varphi_{1} - 36 \epsilon^{2}\alpha^{3}h_{0}^{2}\varphi_{0}\varphi_{1} \, . 
 \end{eqnarray}  

\noindent The equations can be solved in triplets -- at the lowest order ($\ref{eq:O0frI}$), ($\ref{eq:O0dHI}$) and ($\ref{eq:O2sfeI}$) provide a system of dependent equations that can be solved for $\varphi_{0}$, $h_{0}$ -- these are the field equations to lowest order in the field expansion. Then, ($\ref{eq:O1frI}$), ($\ref{eq:O1dHI}$) and ($\ref{eq:O3sfeI}$) correspond to the field equations at next-to-leading order and can be solved for $h_{1}$ and $\varphi_{1}$ etc. At the lowest order we have 

\begin{eqnarray} & & \varphi_{0} = {-1 \pm \sqrt{1 + 8\alpha^{3}} \over 8\alpha^{3} \epsilon} \label{eq:phi0_db1} \\
& & h_{0} =  \dfrac{1 + 2 \alpha^3 \pm \sqrt{1 + 8 \alpha^3}}{6\alpha^3} \, . \label{eq:h0_db1}
\end{eqnarray}

\noindent For $\alpha = 1$, $h_{0} =0$ and $\varphi_{0} = -1/2$ as expected. For $\alpha > 1$, $h_{0}$ is positive for both branches and $h \to 0^{+}$, but if $0 < \alpha < 1$ then $h_{0}$ is negative on one branch and $h \to 0^{-}$. Following the $\varphi_{0} < 0$ branch for which $h_0 \rightarrow 0^{+}$ as $\alpha \rightarrow 1^{+}$, at subsequent orders we have $\varphi_1 = h_1 = h_2 = \varphi_3 = h_3 = 0$, $\varphi_2 = -\tilde{\Lambda}$, $\varphi_4 \neq 0$, and $h_4 \neq 0$. 

In this example, both $h$ and $\varphi$ now possess non-trivial time dependence, but in such a way that $h \varphi' \simeq {\rm constant}$. The Minkowski vacuum state is asymptotically preserved in the sense that $h \to 0$ as $\tau \to \infty$, specifically we have $h = h_{0}\tau^{-1} + {\cal O}(\tau^{-2})$. The behaviour of this solution is similar to the exact, degenerate solution obtained in Ref. \cite{Appleby:2020dko} in the sense that asymptotically $\varphi'' \simeq {\rm constant}$, but there is no exact Minkowski space solution. Still, we have a spacetime that evolves to a low energy state regardless of the presence and magnitude of $\Lambda$. The expansion rate $h(\tau)$ will be sensitive to $\Lambda$, but the Cosmological Constant will only determine how fast the metric evolves to $h \to 0$. Furthermore, $\Lambda$ only enters at lower order in the $\tau$ expansion.

\subsection{Coupling to Ricci Scalar} 
\label{sec:coupling} 

Next we consider a non-minimal coupling to the Ricci scalar which breaks the degeneracy condition. The action and dimensionless field equations read

 \begin{equation} S = \int \sqrt{-g} d^{4}x \left[ {\left( \mpl + M \phi \right) R \over 2} + \epsilon X + {\epsilon^{2} \over \lambda^{3}} X \Box \phi - \lambda^{3}\phi  - \Lambda \right] \end{equation} 

\noindent and 

 \bea 
3 h^2 \left(\Delta ^2+\varphi  \mathcal{M}\right) &=& \tilde{\Lambda }-3 h \varphi ' \left(\mathcal{M}+\epsilon ^2 \left(\varphi '\right)^2\right)+\varphi +\frac{1}{2} \epsilon  \left(\varphi '\right)^2 \label{eq:frIII}\\ 
-2 h' \left(\Delta ^2+\varphi  \mathcal{M}\right) &=&-h \mathcal{M} \varphi '-3 h \epsilon ^2 \left(\varphi '\right)^3+\mathcal{M} \varphi ''+\epsilon ^2 \left(\varphi '\right)^2 \varphi ''+\epsilon  \left(\varphi '\right)^2 \label{eq:dhIII}\\ 
0&=& \varphi '' \left(\epsilon -6 h \epsilon ^2 \varphi '\right) - h^2 \left(6 \mathcal{M}+9 \epsilon ^2 \left(\varphi '\right)^2\right) \nonumber \\
& & \phantom{ggggggggggggg} -3 h' \left(\mathcal{M}+\epsilon ^2 \left(\varphi '\right)^2\right)+3 h \epsilon  \varphi '+1  \label{eq:sfIII} 
\eea 
where $M = \mathcal{M} \lambda$.
 
The system admits a power law expansion of the form of ($\ref{eq:phi_series}$) and ($\ref{eq:h_series}$), with the dominant, leading order terms given by $\varphi_0 = -1/2\epsilon$ and $h_0 = 0$. There is a second solution, with $\varphi_{0} =1/4\epsilon$ and $h_{0} = 1$, but $\varphi$ is generically negative when $\tilde{\Lambda} > 0$ so we do not pursue this branch. Following the branch for which $h_0 = 0$ we have at leading order, $\varphi \sim \tau^2$ and $h \sim \tau^{-3}$. The subdominant terms can be evaluated at progressive orders in $\tau$, the next few are given by $\varphi_1 = h_1 = \varphi_3 = h_3 = 0$, $\varphi_2 = - \tilde{\Lambda} - \mathcal{M}/\epsilon$, $h_2 = \mathcal{M}/3$, $\varphi_4 = - \mathcal{M}^2/9\epsilon$, and $h_4 = -2 \mathcal{M}^2/9$. This supports the existence of an asymptotic Minkowski state despite the departure from degeneracy through the presence of an explicit Ricci coupling. The time dependence of $h$ is sourced by $\mathcal{M}$ at each order in the expansion. 
 
The result of this section -- the existence of an asymptotic vacuum state -- can be generalised to a coupling of the form $\beta^{2-n} \phi^n R$ for some mass scale $\beta$ and constant $n$. In this case, the dimensionless Einstein and scalar field equations read
 \bea 
3 h^2 \left(\Delta ^2 b^n+b^2 \varphi ^n\right) &=& \frac{1}{2} b^n \left(2 \tilde{\Lambda }-6 h \epsilon ^2 \left(\varphi '\right)^3+\epsilon  \left(\varphi '\right)^2\right)+\varphi  b^n-3 b^2 h n \varphi ^{n-1} \varphi ' \label{eq:frV}\\ 
-2 h' \left(\Delta ^2 b^n+b^2 \varphi ^n\right) &=& \epsilon  b^n \left(\varphi '\right)^2 \left(\epsilon  \varphi ''+1\right)+b^2 n \varphi ^{n-1} \varphi '' \nonumber \\
& & \phantom{g} +b^2 (n-1) n \varphi ^{n-2} \left(\varphi '\right)^2 -h \varphi ' \left(3  \epsilon ^2 b^n \left(\varphi '\right)^2+b^2 n \varphi ^{n - 1}\right) \label{eq:dhV}\\ 
0&=& \varphi '' \left(\epsilon -6 h \epsilon ^2 \varphi '\right) - 3 b^{2-n} n \left(2 h^2+h'\right) \varphi ^{n-1} \nonumber \\
& & \phantom{ggggggggggggg} -\left(9 h^2 \epsilon ^2 \left(\varphi '\right)^2+3 \epsilon ^2 h' \left(\varphi '\right)^2-3 h \epsilon  \varphi '-1\right)
  \label{eq:sfV} 
\eea 
where $\beta = b \lambda$. It can be checked that ($\ref{eq:frV}$), ($\ref{eq:dhV}$), and ($\ref{eq:sfV}$) reduce to ($\ref{eq:frIII}$), ($\ref{eq:dhIII}$), and ($\ref{eq:sfIII}$) in the special case $\beta = M$ and $n = 1$. For $n = 2$, we find that the asymptotic series ansatz $\varphi \sim \tau^2$ and $h \sim \tau^{-1}$ (equations ($\ref{eq:phi_series}$) and ($\ref{eq:h_series}$)) solves the system with non-zero coefficients $\varphi_0 \neq -1/2\epsilon$ and $h_0 \neq 0$. This implies the existence of an asymptotic Minkowski state; however, in contrast with the previous case $n = 1$, the asymptotic state is not the well tempered vacuum for which the scalar field evolves according to equation ($\ref{eq:phi_wtvac}$). For $n = 3$, we have confirmed that an asymptotic series ansatz $\varphi \sim \tau$ and $h \sim \tau^{-1}$ solves the system of equations, indicating that an asymptotic Minkowski state exists, although now with different time dependence in the scalar field compared to the well-tempered solution. For general $n$, we conjecture an asymptotic series solution $\varphi \sim \tau^{2/(n - 1)}$ and $h \sim \tau^{-1}$ and note that $n = 3$ is the last case where $\varphi$'s dominant term in the expansion is of integer order. We return to further implications of $n \neq 1$ in the Discussion.

 \subsection{Including Matter}
 \label{sec:matter}
 
 A most natural way to break the degeneracy is with the inclusion of matter fields. In reality, the exact degeneracy equations are never truly satisfied and the dynamics is always `off-shell'\footnote{Following \cite{Charmousis:2011bf}, `on-shell' indicates that the metric is evaluated exactly at the vacuum state and `off-shell' away from it.} due to the presence of dark matter, baryons and radiation. We consider the field equations of the model ($\ref{eq:ac1}$) in the presence of matter:
 
  \bea 
3 \Delta ^2 h^2&=&\varrho + \tilde{\Lambda }-3 h \alpha^3 \epsilon ^2 \left(\varphi '\right)^3+\varphi +\frac{1}{2} \epsilon  \left(\varphi '\right)^2 \label{eq:frIV}\\ 
-2 \Delta ^2 h'&=& \varrho + p -3 h \alpha^3 \epsilon ^2 \left(\varphi '\right)^3+ \alpha^3 \epsilon ^2 \left(\varphi '\right)^2 \varphi ''+\epsilon  \left(\varphi '\right)^2 \label{eq:dhIV}\\ 
0&=& \varphi '' \left(\epsilon -6 h \alpha^3 \epsilon ^2 \varphi '\right) -9 h^2 \alpha^3 \epsilon ^2 \left(\varphi '\right)^2- 3 \alpha^3 \epsilon ^2 h' \left(\varphi '\right)^2+3 h \epsilon  \varphi '+1 \label{eq:sfIV}  \\
0 &=& \varrho ' + 3 h (p+\varrho ) \label{eq:meqIV}
\eea 
where $\rho = \lambda^4 \varrho$ and $P = \lambda^4 p$. Note that equations ($\ref{eq:dhIV}$) and ($\ref{eq:sfIV}$) cannot be equivalent when $\varrho + p \neq 0$, so matter generically breaks the degeneracy of the field equations. We let $\varrho$ represent pressureless dust and fix $p=0$, ignoring any radiation component although its presence would not alter our conclusions. We solve this full system $\left(h, \varphi, \varrho \right)$ beginning from a matter era, i.e., $3 H^2 = \rho$, and show that the state falls to a Minkowski vacuum. This supports the existence of an asymptotic Minkowski state when degeneracy is broken due to the presence of matter. In our simulations, we adopt mass scales, $\tilde{\Lambda} \ll \Delta^2$, and take the matter era initial conditions $h \sim 2/(3\tau)$, $\varphi \sim - \tilde{\Lambda}$, and $3\Delta^2 h^2 \sim \varrho$.

\begin{figure}[h!]
\center
	\subfigure[ ]{
		\includegraphics[width = 0.475 \textwidth]{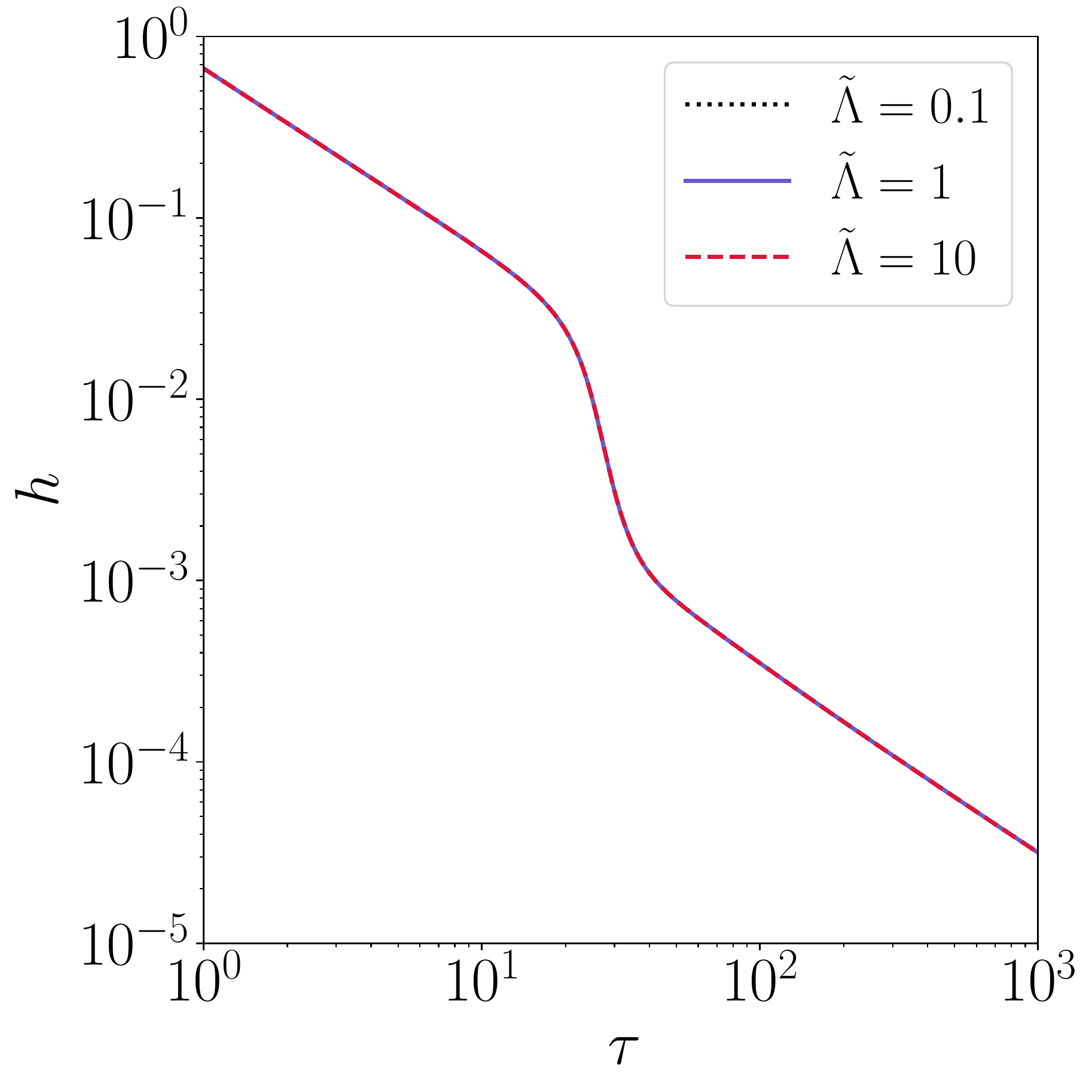}
		}
	\subfigure[ ]{
		\includegraphics[width = 0.475 \textwidth]{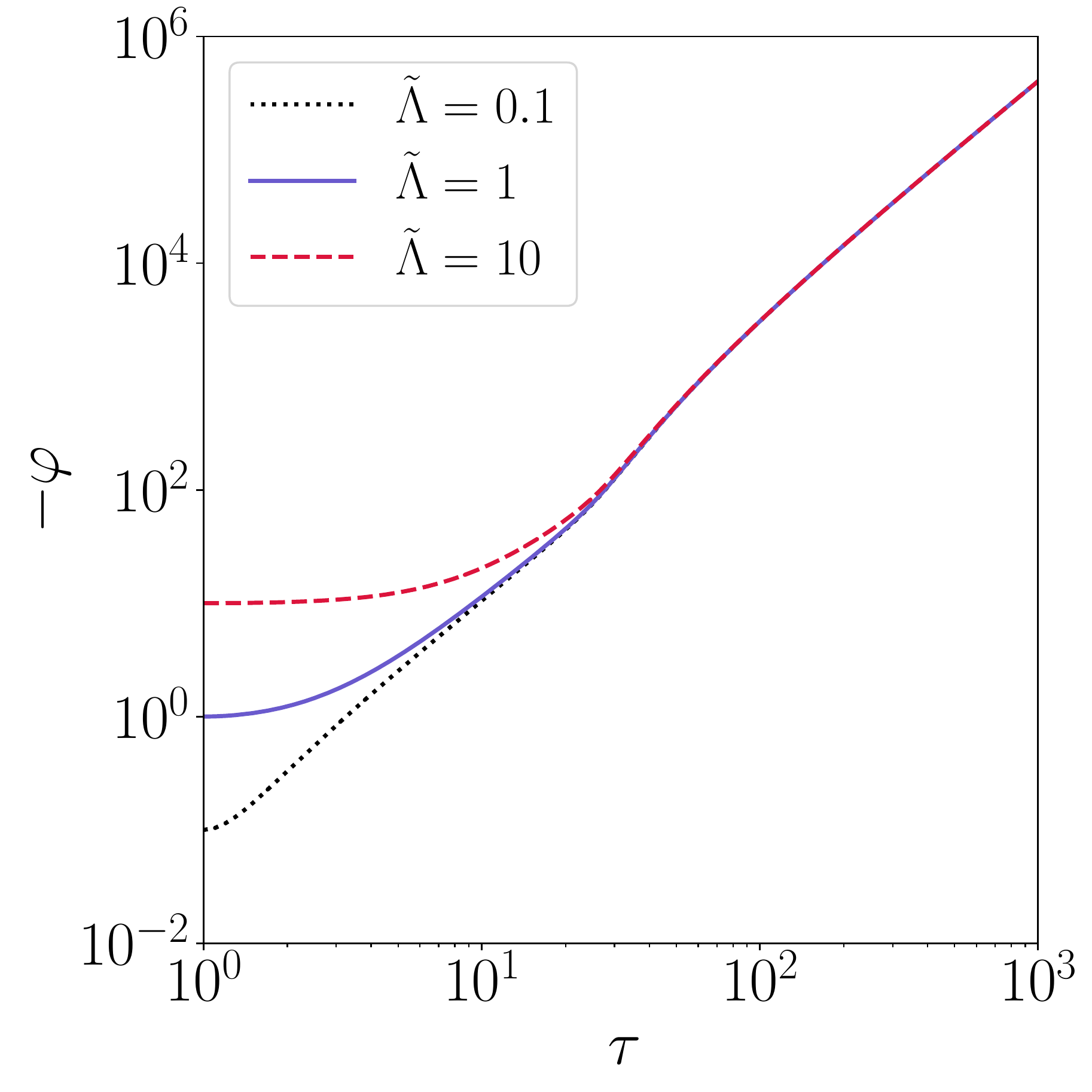}
		}
\caption{Solutions to the coupled gravity-matter equations ($\ref{eq:frIV}$), ($\ref{eq:dhIV}$), ($\ref{eq:sfIV}$), and ($\ref{eq:meqIV}$) with $\tilde{\Lambda} = 10^{0}$ (solid-blue), $\tilde{\Lambda} = 10^{1}$ (medium-dashed-red), $\tilde{\Lambda} = 10^{-1}$ (short-dashed-black), with fixed $\Delta = 10^2$ and $\alpha = 11/10$.}
\label{fig:hphi_db4}
\end{figure}

Figure \ref{fig:hphi_db4} shows numerical solutions to the field equations for various choices of $\tilde{\Lambda}$. Other choices of parameters only lead to similar profiles.

This confirms our assertion of the presence of an asymptotic Minkowski state, which was always reached by the solutions despite beginning from a matter universe. Peculiarly, solutions with $\alpha = 1$ seem to first turn to negative values before transitioning to the degenerate vacuum where the scalar field decelerates, $\ddot{\phi} \sim -\lambda^3/\epsilon$. Admitting a heavier tadpole compared to the braiding, i.e. $\alpha > 1$, resolves this and ensures that $h$ remains positive (cf. Figure \ref{fig:hphi_db4}, left panel). The inclusion of matter preserves the existence of a Minkowski attractor that is approached asymptotically.

\begin{figure}[h!]
\center
\includegraphics[width = 0.475 \textwidth]{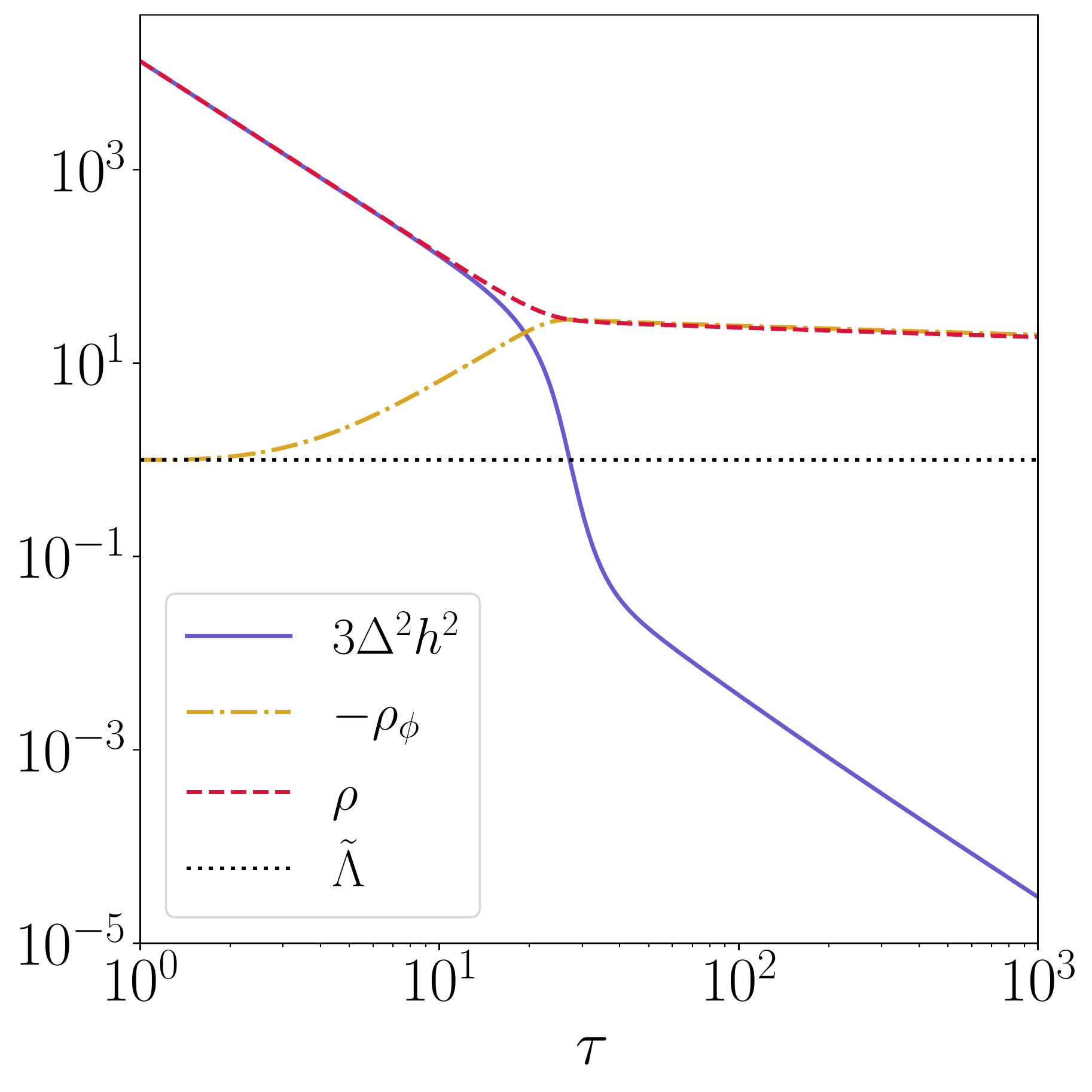}
\caption{Solution to the coupled gravity-matter equations ($\ref{eq:frIV}$), ($\ref{eq:dhIV}$), ($\ref{eq:sfIV}$), and ($\ref{eq:meqIV}$) with $\tilde{\Lambda} = 10^{0}$, $\Delta = 10^2$, and $\alpha = 11/10$.}
\label{fig:rho_db4}
\end{figure}

It is useful to look at how the different densities change during the transition from a matter universe. This is shown in Figure \ref{fig:rho_db4} for $\tilde{\Lambda} / \Delta^2 = 10^{-4}$ and $\alpha = 11/10$. Varying the parameters does not change the overall profile that is presented here. At early times, the Hubble expansion scales with matter as is demanded by the initial conditions. A matter era then persists for a period of time followed by a sharp drop in the expansion rate as the scalar field decelerates and approaches the asymptotic behaviour $\varphi \sim \tau^{2}$, $h \sim \tau^{-1}$. 

At no point have we introduced any unusual fine tuning of mass parameters. The matter contribution is nonzero throughout in Figure \ref{fig:rho_db4}. The Minkowski state is approached regardless of any degeneracy condition being satisfied, despite the presence of an arbitrary vacuum energy and the presence of matter. However, since the model approaches a Minkowski space vacuum state asymptotically, we do not expect it to provide a viable cosmic history. One might hope that alternative methods of breaking the degeneracy condition might impact the expansion rate $h$ such that approximate pseudo-de Sitter might be asymptotically realised. We provide one such example in the following section.  

\subsection{Introducing Scalar Field Mass}
\label{sec:5}

If we introduce a mass term for the Galileon, the action and cosmological field equations (in dimensionless units) are 

\begin{equation} S = \int \sqrt{-g} d^{4}x \left[ {M_{\rm pl}^{2} R \over 2} + \epsilon X + {\epsilon^{2} \over \lambda^{3}} X \Box \phi - \lambda^{3}\phi - {m_{\phi}^{2} \over 2}\phi^{2} - \Lambda \right] \end{equation} 

\noindent and 
 \bea 
3\Delta^{2} h^2 &=& \tilde{\Lambda} + {\epsilon \over 2} (\varphi')^{2} - 3 \epsilon^{2} h(\varphi')^{3}   + \varphi + {\tilde{m}_{\phi}^{2} \over 2} \varphi^{2} \label{eq:frII}\\ 
-2\Delta^{2} \, h' &=&
 \epsilon^{2} (\varphi')^{2} \varphi'' - 3 \epsilon^{2}  h (\varphi')^{3}
+ \epsilon (\varphi')^{2}  \label{eq:dhII}\\ 
0&=& \varphi''\,\left[\epsilon - 6\epsilon^{2} h\varphi'\right]+3\epsilon h\varphi' + 1 - 3\epsilon^{2}  (\varphi')^{2} (h'+3h^2) + \tilde{m}_{\phi}^{2}\varphi   \label{eq:sfII} 
\eea 

\noindent where $\tilde{m}_{\phi} = m_{\phi}/\lambda$. We have fixed $\alpha = 1$, so the Galileon has an exact $h=0$ solution when $\tilde{m}_{\phi} = 0$.  

This system of equations is non-linear and has multiple scales, so it is very difficult to make any general statements. We take the following mass scales $\tilde{\Lambda} \sim {\cal O}(1)$, $\tilde{m}_{\phi} \sim {\cal O}(1)$ and $\Delta \gg 1$. With this choice, the two mass scales associated with the field $\phi$ -- $\lambda$ and $m_{\phi}$ -- are of the same magnitude, which in turn are of the same order as the Cosmological Constant (which is arbitrary, so far). All mass scales are smaller than the Planck mass, which dictates the size of $\Delta$. At least, with this choice we are not fine tuning any mass scales in the action and are working in a sub-Planckian regime. Note that the degenerate vacuum solution obtained in Ref. \cite{Appleby:2020dko} does not rely on $\lambda$ being of the same order of magnitude as $\Lambda$, so we could introduce a hierarchy such that $\Delta \gg \tilde{\Lambda} \gg 1$ but in the absence of any reason to, we do not introduce any hierarchy between the mass scale associated with the field $\phi$ and $\Lambda$. 

We use the Friedmann equation to fix as an initial condition $\varphi_{i} = \varphi'_{i} = 0$ and

\begin{equation}\label{eq:hi} h_{i}^{2} = {\tilde{\Lambda} \over 3 \Delta^{2}} \ll 1 \end{equation}

\noindent which is the standard general relativity vacuum state. This would be an exact solution if $\lambda = 0$.  Under these conditions we can obtain an approximate solution to these equations using the fact that $\Delta \gg 1$ and $h \ll 1$. Anticipating that in this regime $h$ is slowly rolling, the scalar field equation can be approximated as 

\begin{equation} \epsilon \varphi'' + 3 \epsilon h_{i} \varphi' + \tilde{m}_{\phi}^{2}\varphi \simeq -1 \, . \end{equation} 

\noindent Using the initial conditions $\varphi_{i} = \varphi'_{i} = 0$, we have as a leading order approximation 

\begin{equation} \varphi \simeq {1 \over \tilde{m}_{\phi}^{2}} \left[  e^{-3\epsilon h_{i}\tau/2} \cos \left({\tilde{m}_{\varphi} \tau \over \sqrt{\epsilon}}\right) - 1  \right] \end{equation} \, .

\begin{figure}[h!]
\center
	\subfigure[ ]{
		\includegraphics[width = 0.475 \textwidth]{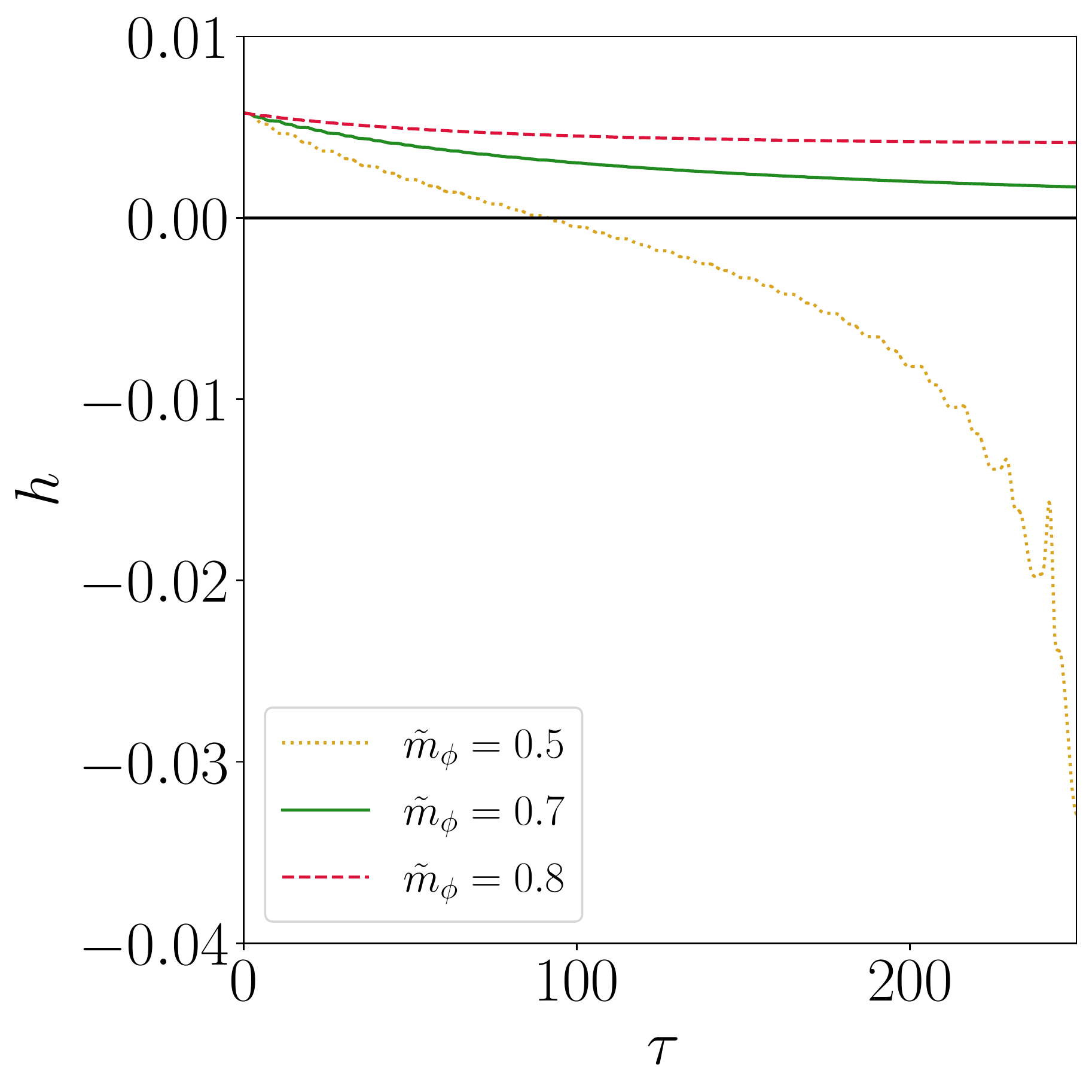}
		}
	\subfigure[ ]{
		\includegraphics[width = 0.475 \textwidth]{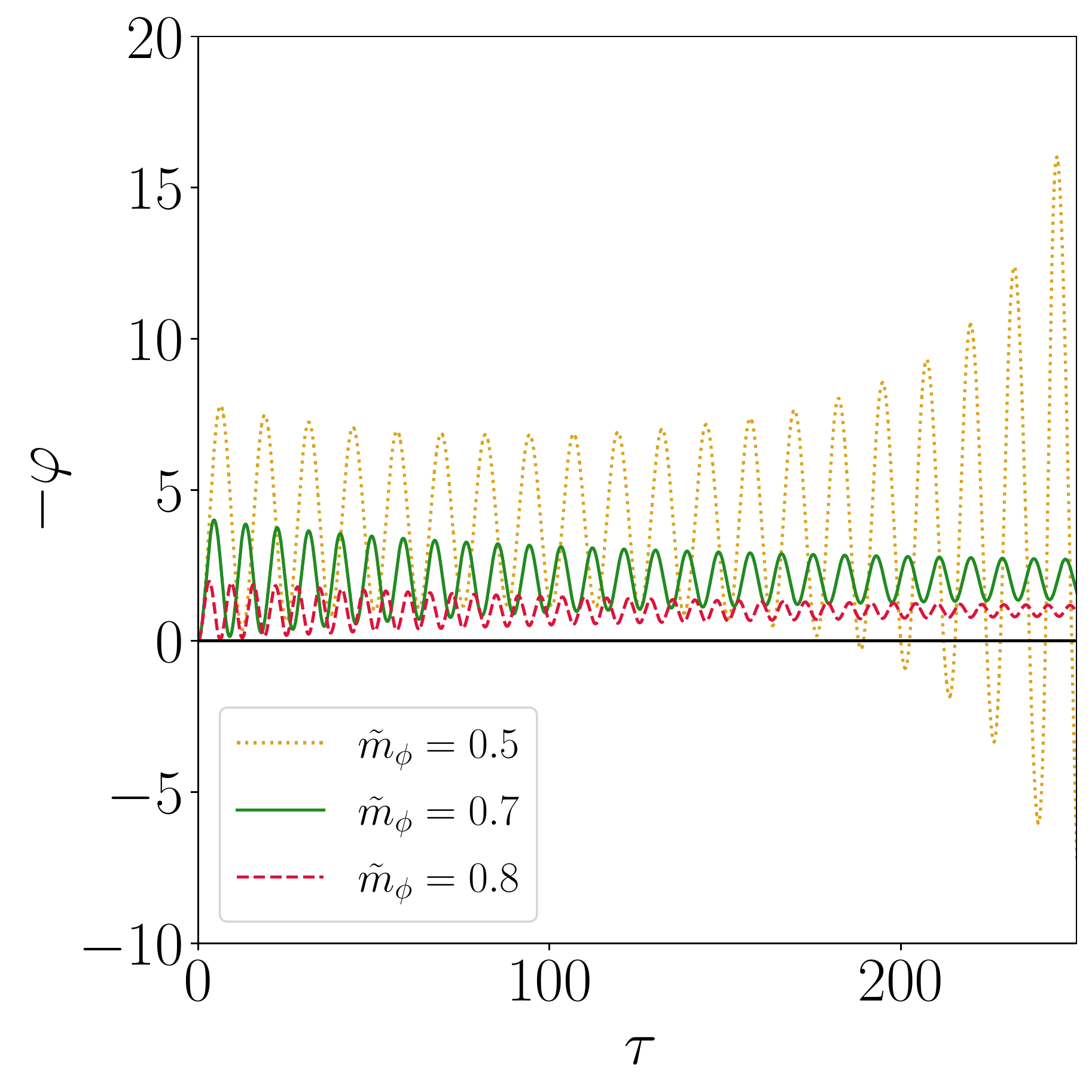}
		}
\caption{Numerical solutions to equations ($\ref{eq:frII}$), ($\ref{eq:dhII}$), and ($\ref{eq:sfII}$) with $\tilde{m}_{\phi} = 0.5$ (dotted gold), $\tilde{m}_{\phi} = 0.7$ (solid green), $\tilde{m}_{\phi} = 0.8$ (dashed red), fixing $\tilde{m}_\phi = 1$, and $\Delta = 10^2$.}
\label{fig:ds_db2}
\end{figure}

 \noindent In turn, the expansion rate $h$ has an oscillating component, sourced by $\varphi$. Note that in this model, the scalar field exhibits oscillatory behaviour and hence is potentially bounded. This behavior is supported by our numerical solutions, provided the mass scales satisfy $\tilde{\Lambda} \sim {\cal O}(\tilde{m}_{\phi})$ and $\Delta \gg 1$. Some examples are shown in Figure \ref{fig:ds_db2}, where we fix $\Delta = 10^{2}$, $\tilde{\Lambda} = 1$, $\epsilon = 1$ and $\tilde{m}_{\phi} = 0.8, 0.7, 0.5$ (cf. red-dashed, green-solid and gold-dotted lines, respectively) then proceed to numerically evolve the dynamical system with initial conditions $\varphi_{i} = \varphi' = 0$ and $h_{i}^{2} = \tilde{\Lambda}/(3\Delta^{2})$. For suitable initial conditions and mass scales, the scalar field undergoes damped oscillations, and the expansion rate $h$ freezes to an asymptotically frozen, mildly oscillating phase (cf. red-dashed lines).

However, when $\tilde{m}_{\phi}$ is sufficiently low compared to the vacuum energy, i.e., $\tilde{\Lambda} > \tilde{m}_\phi$, the solution becomes unstable, with $h(\tau)$ descending to negative values and the scalar field amplitude growing rapidly. This is shown in Figure \ref{fig:ds_db2} (cf. dotted gold lines, $\tilde{m}_{\phi} = 0.5$).

The dynamics of $h$ is tied to the envelope of $\varphi$. Both $h$ and the envelope of $\varphi$ are initially decreasing functions of $\tau$, and the relative rate at which they decrease is important. If $h$ approaches zero while the amplitude of $\varphi$ remains sufficiently large, then $h$ crosses zero and continues catastrophically towards arbitrary negative values. It is at the $h=0$ crossing that the scalar field envelope starts to grow. However, if the amplitude of $\varphi$ decays sufficiently quickly, then $h$ becomes frozen to an approximately constant value. Neither of the fields $h$ or $\varphi$ are actually constant, they continue to oscillate with decreasing amplitude. Still, $h$ can become frozen into an apparent de Sitter like state.

\begin{figure}[h!]
\center
	\includegraphics[width = 0.475 \textwidth]{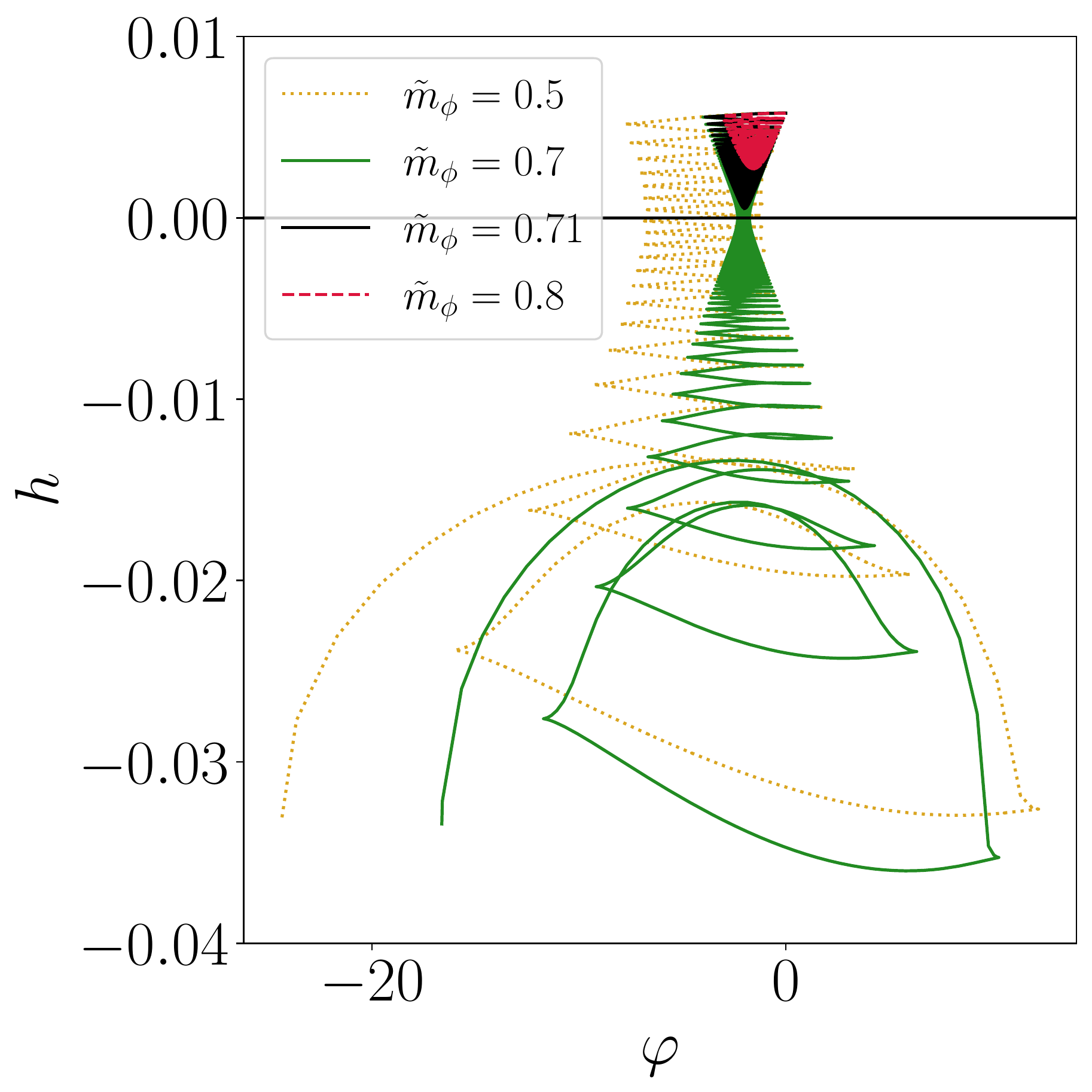}
\caption{$h$ as a function of $\varphi$ for the system ($\ref{eq:frII}$), ($\ref{eq:dhII}$), and ($\ref{eq:sfII}$) with $\tilde{\Lambda} = 1$, $\Delta = 10^2$ and varying $\tilde{m}_{\phi}$. If the mass of the scalar field is sufficiently low, the expansion rate $h$ evolves beyond $h=0$ to arbitrarily negative values. Above a particular threshold value of $\tilde{m}_{\phi}$, the expansion rate freezes to an approximately constant state (cf. black-solid, red-dashed lines).}
\label{fig:h_varphi}
\end{figure}

In Figure \ref{fig:h_varphi} we present $h$ as a function of $\varphi$ for different values of $\tilde{m}_{\phi}$. For each dynamical track, we fix the same initial conditions $\varphi_{i} = \varphi'_{i} = 0$ and $h_{i}$ given by ($\ref{eq:hi}$), fix $\tilde{\Lambda} = 1$, $\Delta = 10^{2}$, $\epsilon = 1$ and allow $\tilde{m}_{\phi}$ to vary over the range $0.5 < \tilde{m}_{\phi} < 0.8$. If the scalar field mass is sufficiently low, then $h$ evolves beyond $h=0$ and to negative values, and the envelope of the $\varphi$ oscillations grows (cf. yellow-dashed, green-solid lines). If $\tilde{m}_{\phi}$ is large, then the envelope of $\varphi$ oscillations decays more rapidly and $h$ becomes frozen to an approximately constant value (cf. black-solid, red-dashed tracks). 

Unfortunately, for this state to mimic the observed late-time accelerating epoch of the Universe, the parameters in the model should be fine tuned. In particular the mass of the scalar field must take a precise value -- too large and $h$ would freeze to an unacceptably large value and too small and $h$ would slide catastrophically to arbitrary negative values. Still, the dynamics of this particular case possess some welcome features -- there is a range of parameter values for which the scalar field and $h$ are bounded, and $h$ can freeze to a dynamical low energy state. There is no longer a stable $h=0$ Minkowski space solution, however. Counter to the previous examples in this section, the presence of a mass term eliminates the possibility of a flat spacetime solution, even asymptotically.

We acknowledge that the question as to whether these models inherently require fine tuning remains open. In this section, we have provided a simple example for which the fields asymptotically freeze to `constant' values (although there is no exact, constant solution, and the fields continue to evolve indefinitely). The end point of the dynamics in this class of models depends on some combination of the $(R, \phi, X, \Box \phi)$-functional dependence of the Lagrangian, coupling constants and the initial conditions of the fields. In contrast, the end-point in the standard model is generically the algebraic relation $3H^{2} = 8\pi G \Lambda$, making fine tuning inevitable. Although the simple example provided in this section has its own stability issues, the dynamical nature of the solution opens up the possibility of ameliorating fine tuning, as the fields $h$, $\varphi$ can approximately freeze and their asymptotic values become a function of the initial conditions, age/dynamics of the Universe in addition to the coupling constants in the action.

Hence, this class of models rephrase the Cosmological Constant problem in a novel way. It remains an open question whether a `natural' end point to the dynamics of the expansion rate $h$ can be realised. The current work constitutes an interesting step forward in this regard, as we found that the strict enforcement of the degeneracy condition is not required. This means that we do not need to impose any exact conditions on the coupling constants of the theory, removing this particular element of fine tuning from the proposal. The idea that fields can `freeze' into approximately constant values which depend on the prior dynamics of the system is made possible with asymptotic degenerate states. These are not constant attractor solutions predictable solely from coupling constants of the theory.

To end, we comment that a Vainshtein mechanism sourced by the braiding $\left(\mathcal{L}\sim \left( \partial \phi \right)^2 \Box \phi\right)$ can be expected to settle in at small distances where nonlinearities become the dominant contributions to the equations of motion. These scales are irrelevant to the current discussion, but may nonetheless be considered in a different work which looks at spherically symmetric solutions in degenerate and non-degenerate models.

 \section{Discussion}
 \label{sec:6}
 
The existence of an exact Minkowski solution despite the presence of an arbitrary vacuum energy for the class of models studied in this work requires a degeneracy relation, which must be solved exactly \cite{Appleby:2020dko, Bernardo:2021bsg}. In this work we have shown that even when it is not, we do not expect the standard relation $3M_{\rm pl}^{2} H^{2} = \Lambda$, $\phi = {\rm constant}$ to necessarily be a solution to the dynamical system when the tadpole is present. Rather, when the degeneracy equation is broken the metric will also be time dependent in tandem with the scalar field. The Minkowski solution is a dynamical attractor when the degeneracy condition holds exactly, and we have found that it is also an attractor without imposing any exact relation between terms in the Lagrangian. We have considered some simple ways in which the degeneracy condition can be broken, and studied the dynamics that result when both the metric and scalar field are free to evolve.

We confirmed that departures from degeneracy forces the metric into a dynamical state, but nonetheless for most cases an asymptotic Minkowski solution was retained. Unsurprisingly, this may not be precisely the well tempered vacuum, particularly in cases where the shift symmetry is broken in the Einstein equation. The functional time dependence of the solution is related to presence or absence of shift symmetry that the degenerate model possesses. The model in section \ref{sec:unequal_mass} and the linear coupling $\sim \phi R$ in section \ref{sec:coupling} preserve the symmetry $\phi \to \phi +c$ and hence exhibit $\varphi \sim \tau^{2}$ behaviour on approach to the Minkowski asymptote. Adding a non-linear coupling $\sim \phi^{n}R$ or mass term $m_{\phi}^{2}\phi^{2}$ changes mass dimension operators in the action and correspondingly the time dependence of the scalar field. Degeneracy breaking with a mass term can produce a low energy de Sitter phase which could potentially reconcile the background evolution with observations. Unfortunately, the solution faces something of a cliff edge towards an unstable $h < 0$ state. Admittedly, this may be considered as another kind of fine tuning, but it opens the possibility of having a consistent cosmology independent of a large Cosmological Constant. Overall, our results broaden the horizon for what could be classified as self-tuning models and the tadpole plays a major role. This generic dynamical behaviour has been similarly used in Ref. \cite{Khan:2022bxs}, which found asymptotic de Sitter solutions without a degeneracy condition being imposed. It would be of interest to study the generality of Ref. \cite{Khan:2022bxs} and consider how common asymptotic de Sitter states are.

Throughout this work we have used `shift symmetry' as a loosely defined label. A conservative shift symmetric model is made up of only derivatives of the scalar field in the Lagrangian such that one can define a Noether current $J^\alpha \left[ \partial \phi\left(x\right) \right]$ satisfying a conservation law $ \nabla_\alpha J^\alpha = 0 $ which corresponds to the scalar field equation. In the presence of a tadpole, one may instead write down $\nabla_\alpha \left( J^\alpha - \lambda^3 x^\alpha \right) = 0$, identifying a conserved charge $Q = J^t - \lambda^3 t$ which is related to the Noether charge by a time reparametrization. This manifests at the level of the action, or the field equations, upon the application of a shift transformation which only artificially influences the dynamics through the initial conditions entering the Hamiltonian constraint at an arbitrary time. In this way, we regard the exactly degenerate model in section \ref{sec:3} to be shift symmetric `on-shell' and its vacuum state is insensitive to the value of the Cosmological Constant. In the nondegenerate models of section \ref{sec:4}, and similarly the model in section \ref{sec:3} `off-shell', the dynamics will depend subdominantly on the Cosmological Constant as the system asymptotes to the degenerate vacuum. The mass term and nonlinear conformal couplings are also considered as shift symmetry breaking in this regard, as they alter the dynamical behaviour on approach to the degenerate solution.

We emphasize that the models considered here must be further studied to ascertain whether they could indeed represent viable cosmologies. For one, viable models should not only have a consistent background, but also possess only healthy perturbations on top of it. The degenerate model (section \ref{sec:3}), when evaluated `on-shell', is ghost-free but suffers from a Laplace instability after a time $t \gtrsim \left( M_\text{pl}/\lambda^3 \right)^{1/2}$ \cite{Appleby:2020dko}. The models considered in this work cannot be evaluated on an exact static background, but asymptotically we can determine their stability. For example, for the model in which we relaxed the mass scales (section \ref{sec:unequal_mass}), once the fields $H$ and $\phi$ have relaxed to their $t \gg \lambda^{-1}$ asymptotic forms $\phi \sim \lambda^{3} t^{2}$ and $H \sim t^{-1}$, we can deduce that after a time $t \gtrsim \left( M_\text{pl} \kappa^{3/2}/\lambda^{9/2} \right)^{1/2}$ in this state the scalar perturbations of the dynamical fields will similarly exhibit Laplace instability. A lighter tadpole compared to the braiding can keep the Laplace stability in check while the system asymptotically evolves to the Minkowski vacuum for a longer period, but never indefinitely. This, among other effects at linear cosmological perturbations, can be studied conveniently using the effective field theory formalism (Appendix \ref{sec:eft}). Studying the field equations in the regime in which the Laplace condition is violated is an interesting open problem. 

Other future directions may be considered. One, it would be interesting to see if similar departures from the Fab Four can be obtained. Second, models containing light dynamical fields must suppress fifth forces to satisfy Solar system constraints and be considered viable. Screening mechanisms are a potential loophole, but the question of whether these models exhibit this nonlinear feature is unresolved. Given that the presence of matter breaks the degeneracy condition, a `static' metric is unlikely to solve the coupled scalar-Einstein field equations in the presence of a central mass. Hence what spacetime replaces the role of the Schwarzschild metric remains to be found. The lack of cosmological constraints on degenerate and (now) nearly-degenerate models must also be given attention, once a viable mechanism to generate late-time accelerated expansion has been introduced. Finally, studying the conditions under which all fields remain bounded could help to reconcile the self tuning mechanism with high energy physics.

\section*{Acknowledgements}
The authors would like to thank Eric Linder, Arnaz Khan, and Andy Taylor for helpful suggestions and discussions. SAA is supported by an appointment to the JRG Program at the APCTP through the Science and Technology Promotion Fund and Lottery Fund of the Korean Government, and was also supported by the Korean Local Governments in Gyeongsangbuk-do Province and Pohang City.

\appendix

\section{Linear perturbations and effective field theory functions}
\label{sec:eft}

The behavior of linear perturbations of the action ($\ref{eq:action}$) can be conveniently studied through three effective field theory functions \cite{Creminelli:2008wc, Gubitosi:2012hu, Bloomfield:2012ff, Gleyzes:2013ooa, Bellini:2014fua}: the kineticity ($\alpha_K$), braiding ($\alpha_B$), and the running mass ($\alpha_M$). These are explicitly given in terms of the potentials by \cite{Bellini:2014fua}
\begin{equation}
\label{eq:kineticity}
    \alpha_K = \dfrac{2X\left(K_X + 2XK_{XX}\right) + 12H \dot{\phi} X \left( G_{3X} + X G_{3XX} \right)}{H^2 \left( \mpl + F \right)} \, ,
\end{equation}
\begin{equation}
\label{eq:braiding}
    \alpha_B = \dfrac{\dot{\phi} \left( 2 X G_{3X} - F_\phi \right)}{H \left( \mpl + F \right)} \, ,
\end{equation}
and
\begin{equation}
\label{eq:running_mass}
    \alpha_M = \dfrac{F_\phi \dot{\phi}}{H \left( \mpl + F \right)} \, .
\end{equation}
The absence of ghost and Laplace instabilities (in the scalar sector) can then be ensured by the conditions
\begin{equation}
\label{eq:no_ghost}
    D = \alpha_K + \dfrac{3}{2} \alpha_B^2 \geq 0
\end{equation}
and
\begin{equation}
\label{eq:no_gradient}
    c_s^2 = \dfrac{\left( 1 - (\alpha_B/2) \right)}{D} \left( \alpha_B + 2 \alpha_M \right) + \dfrac{ \partial_t \left( H \alpha_B \right) }{ H^2 D } - \dfrac{2 \dot{H}}{H^2 D} \geq 0
\end{equation}
where the former guarantees the correct sign in the kinetic term while the latter keeps the sound speed squared positive. The quadratic action for the scalar ($\zeta$) and tensor modes ($h_{ij}$) can be written as
\begin{equation}
    S^{(2)} = \int dt d^3x \ a^3 \left[ \dfrac{2 \left( \mpl + F \right)D}{\left(2 - \alpha_B \right)^2} \left( \dot{\zeta}^2 - \dfrac{c_s^2}{a^2} \left( \partial_i \zeta \right)^2 \right) + \dfrac{\left(\mpl + F \right)}{8} \left( \dot{h}_{jk}^2 - \dfrac{1}{a^2} \left( \partial_i h_{jk} \right)^2 \right) \right] \, .
\end{equation}
This also reveals that in action ($\ref{eq:action}$), the tensor modes propagate on the light cone.

As an example, substituting $K(X) = \epsilon X$, $G(X) = -\epsilon^2 X / \kappa^3$, and $F(\phi) = \beta^{2 - n} \phi^n$, covering the models in section \ref{sec:4}, we obtain
\begin{equation}
    \alpha_K = \epsilon \dot{\phi^2} \dfrac{ 1 - \left( 6 \epsilon H \dot{\phi} / \kappa^3 \right) }{H^2 \left( \mpl + \beta^{2 - n} \phi^n \right)} \, ,
\end{equation}
\begin{equation}
    \alpha_B = - \dfrac{ \dot{\phi} \left( \left( \epsilon^2 \dot{\phi}^2 / \kappa^3 \right) + n \beta^{2 - n} \phi^{n - 1} \right) }{ H \left( \mpl + \beta^{2 - n} \phi^n \right) } \, ,
\end{equation}
and
\begin{equation}
    \alpha_M = \dfrac{n \beta^{2 - n} \phi^{n - 1} \dot{\phi}}{H \left( \mpl + \beta^{2 - n} \phi^n \right)} \, .
\end{equation}
Neither the scalar potential $V\left(\phi\right)$ nor the tadpole appears in the above expressions. The stability of the scalar perturbations can be assessed using ($\ref{eq:no_ghost}$) and ($\ref{eq:no_gradient}$).

%\bibliographystyle{JHEP}
%\bibliography{refs}

\providecommand{\noopsort}[1]{}\providecommand{\singleletter}[1]{#1}%

\providecommand{\href}[2]{#2}\begingroup\raggedright\endgroup

\end{document}